\documentclass[aps,prb,reprint,amsfonts,amsmath,amssymb,showpacs,groupedaddress,superscriptaddress]{revtex4-1}
\usepackage{bm}
\usepackage{color}
\usepackage{graphicx}
\usepackage[colorlinks,urlcolor=blue,linkcolor=blue,anchorcolor=blue,citecolor=blue,bookmarks]{hyperref}

\begin{document}

\title{Comprehensive study of the global phase diagram in the triangular $J$-$K$-$\Gamma$ model}

\author{Shi Wang}
\affiliation{National Laboratory of Solid State Microstructures and School of Physics, Nanjing University, Nanjing 210093, China}

\author{Zhongyuan Qi}
\affiliation{College of Physics Science and Technology, Yangzhou University, Yangzhou 225002, China}

\author{Bin Xi}
\affiliation{College of Physics Science and Technology, Yangzhou University, Yangzhou 225002, China}

\author{Wei Wang}
\affiliation{School of Science, Nanjing University of Posts and Telecommunications (NUPT), Nanjing 210023, China}

\author{Shun-Li Yu}
\email{slyu@nju.edu.cn}
\affiliation{National Laboratory of Solid State Microstructures and School of Physics, Nanjing University, Nanjing 210093, China}
\affiliation{Collaborative Innovation Center of Advanced Microstructures, Nanjing University, Nanjing 210093, China}

\author{Jian-Xin Li}
\email{jxli@nju.edu.cn}
\affiliation{National Laboratory of Solid State Microstructures and School of Physics, Nanjing University, Nanjing 210093, China}
\affiliation{Collaborative Innovation Center of Advanced Microstructures, Nanjing University, Nanjing 210093, China}

\date{\today}

\begin{abstract}
The celebrated Kitaev honeycomb model provides an analytically tractable example with an exact quantum spin liquid ground state. While in real materials, other types of interactions besides the Kitaev coupling ($K$) are present, such as the Heisenberg ($J$) and symmetric off-diagonal ($\Gamma$) terms, and these interactions can also be generalized to a triangular lattice. Here, we carry out a comprehensive study of the $J$-$K$-$\Gamma$ model on the triangular lattice covering the full parameters region, using the combination of the exact diagonalization, classical Monte Carlo and analytic methods, with an emphasis on the effects of the $\Gamma$ term. In the HK limit ($\Gamma=0$), we find five quantum phases which are quite similar to their classical counterparts. Among them, the stripe-A and dual N\'{e}el phase are robust against the introduction of the $\Gamma$ term, in particular the stripe-A extends to the region connecting the $K=-1$ and $K=1$ for $\Gamma<0$. Though the 120$^\circ$ N\'{e}el phase also extends to a finite $\Gamma$, its region has been largely reduced compared to the previous classical result. Interestingly, the ferromagnetic (dubbed as FM-A) phase and the stripe-B phase are unstable in response to an infinitesimal $\Gamma$ interaction. Moreover, we find five new phases for $\Gamma\ne 0$ which are elaborated by both the quantum and classical numerical methods. Part of the space previously identified as 120$^\circ$ N\'{e}el phase in the classical study is found to give way to the modulated stripe phase. Depending on the sign of the $\Gamma$ term, the FM-A phase transits into the FM-B ($\Gamma>0$) and FM-C ($\Gamma<0$) phase with different spin orientations. Similarly, the stripe-B phase transits into the stripe-C ($\Gamma>0$) and stripe-A ($\Gamma<0$). Around the positive $\Gamma$ point, due to the interplay of the Heisenberg, Kiatev and $\Gamma$ interactions, we find a possible quantum spin liquid in a noticeable region with a continuum in spin excitations.

% Near the antiferromagetic Kitaev point, we show that the stripe order is selected out of the classic nematic order via the order-by-disorder mechanism. Also, the order-by-disorder mechanism selects the ferromagnetic ordered states as the ground state out of the degenerate classical states for the pure $\Gamma$ models.
% We find that the quantum phase diagram is quite similar to its classical counterpart in the large region, suggesting that the Kitaev physics would have emerged largely in the classical limit.
\end{abstract}

\maketitle

\section{\label{sec:Introduction}Introduction}

Geometric frustration, which arises when the lattice geometry gives rise to constraints that not every exchange bond can be simultaneously minimized in energy, plays an important role in various kinds of magnetic systems. The nearest-neighbor (NN) antiferromagetic (AFM) Heisenberg model on the triangular lattice is a typical example, once two of the spins on an elementary triangle are antiparallel to satisfy their antiferromagnetic interaction, the third one can no longer point in a direction opposite to both other spins. In particular, the spin-1/2 case has attracted numerous interests and was extensively studied after the seminal prediction by Anderson of a ``quantum spin liquid" (QSL) where a strong quantum fluctuation prevents any long-range order down to the zero temperature~\cite{Anderson1973}, although it has been shown by later studies that the ground state (GS) have a classical magnetic order with each spin on a triangle pointing to 120$^\circ$ with respect to each other~\cite{PhysRevLett.99.127004,PhysRevLett.82.3899,PhysRevB.50.10048,PhysRevLett.60.2531}. When the interactions beyond the NN Heisenberg type are included which introduce further frustration, the system has a much richer phase diagram including the 120$^\circ$ N\'{e}el state, stripe states and QSL states~\cite{PhysRevB.91.014426,PhysRevB.92.041105,PhysRevB.92.140403,PhysRevB.96.165141,PhysRevB.93.144411,JPSJ.83.093707,PhysRevB.94.121111,PhysRevB.96.075116,PhysRevLett.120.207203,PhysRevB.95.165110,PhysRevLett.123.207203,PhysRevX.9.031026,wu2020exact}. All these studies have revealed that geometric frustated systems show quite different behavior from that of the non-frustated system.

On the other hand, exchange frustation in systems with strongly anisotropic magnetic interactions has been shown to be another promising approach to explore exotic quantum spin states. Like geometric frustation, the effect of exchange frustration is to prevent the formation of long range magnetic order and give raise to a residual ground-state entropy. The spin-$1/2$ Kitaev model~\cite{Kitaev2006} on honeycomb lattice, which has both gapped and gapless QSL states supporting fractionalized excitations, is a celebrated example of a model with exchange frustration. In this model, the spins subject to the Kitaev interactions consisting of nearest neighbor Ising-type interactions, with the quantization axis depending on the spatial orientation of an exchange bond. Because of its theoretical importance and potential application in quantum computing, great efforts have been made to search for a solid-state realization of the Kitaev model. G. Khaliullin \emph{et al.,}~\cite{Khaliullin2005,PhysRevLett.102.017205} proposed that this highly anisotropic Kitaev interaction can be realized in 4d/5d systems with a low spin state of $d^5$ configuration, such as iridates A$_2$IrO$_3$ (A = Na, Li). In these systems, the bond-directional interactions originate from the joint effects of strong spin-orbital coupling (SOC), electron interactions, $d^5$ configuration and 90$^\circ$ bond geometry formed by edge sharing octahedra. However, in real materials, other types of interaction besides the Kitaev coupling are present, and these interactions may induce other interesting ordered and disordered phases. The simplest extension of the pure Kitaev model is the Heisenberg-Kitaev (HK) model~\cite{PhysRevLett.105.027204,10.1146/annurev-conmatphys-033117-053934}, in which the NN Heisenberg interaction is also taken into account. This model has been extensively studied by various numerical methods~\cite{PhysRevLett.110.097204,PhysRevB.83.245104,PhysRevB.84.100406,PhysRevB.90.195102,PhysRevLett.119.157203}, which reveal the presence of four magnetically ordered phases with collinear spin patterns of ferromagnetic (FM), AFM, stripe, and zigzag types, besides extended spin-liquid phases near the Kitaev limits. Considering the most idealized crystal structure, another interaction beyond the HK model also must be included, i.e. bond dependent symmetric off-diagonal exchange, which is called the $\Gamma$ interaction~\cite{PhysRevLett.112.077204,PhysRevB.93.214431,PhysRevB.96.115103}. Thus, the generic NN exchange Hamiltonian for the undistorted hexagonal compounds is the $J$-$K$-$\Gamma$ model, where the Heisenberg ($J$), Kitaev ($K$) and $\Gamma$ interactions are all included. Finite $\Gamma$ further enriches the phase diagram by adding non-collinear and incommensurate spiral phases~\cite{PhysRevLett.112.077204,Winter_2017,10.1038/s42254-019-0038-2}. Moreover, in real materials, such as A$_2$IrO$_3$ and $\alpha$-RuCl$_{3}$, the dominant interactions are the $\Gamma$ and FM Kitaev terms which originate from both direct $d$-$d$ and anion mediated $d$-$p$ electron transfer, while the Heisenberg term has the smallest strength since it predominantly originates from the weak direct $d$-$d$ electron transfer~\cite{PhysRevLett.112.077204,PhysRevB.93.214431,PhysRevB.96.115103}.

\begin{figure}
    \centering
    \includegraphics[width=\columnwidth]{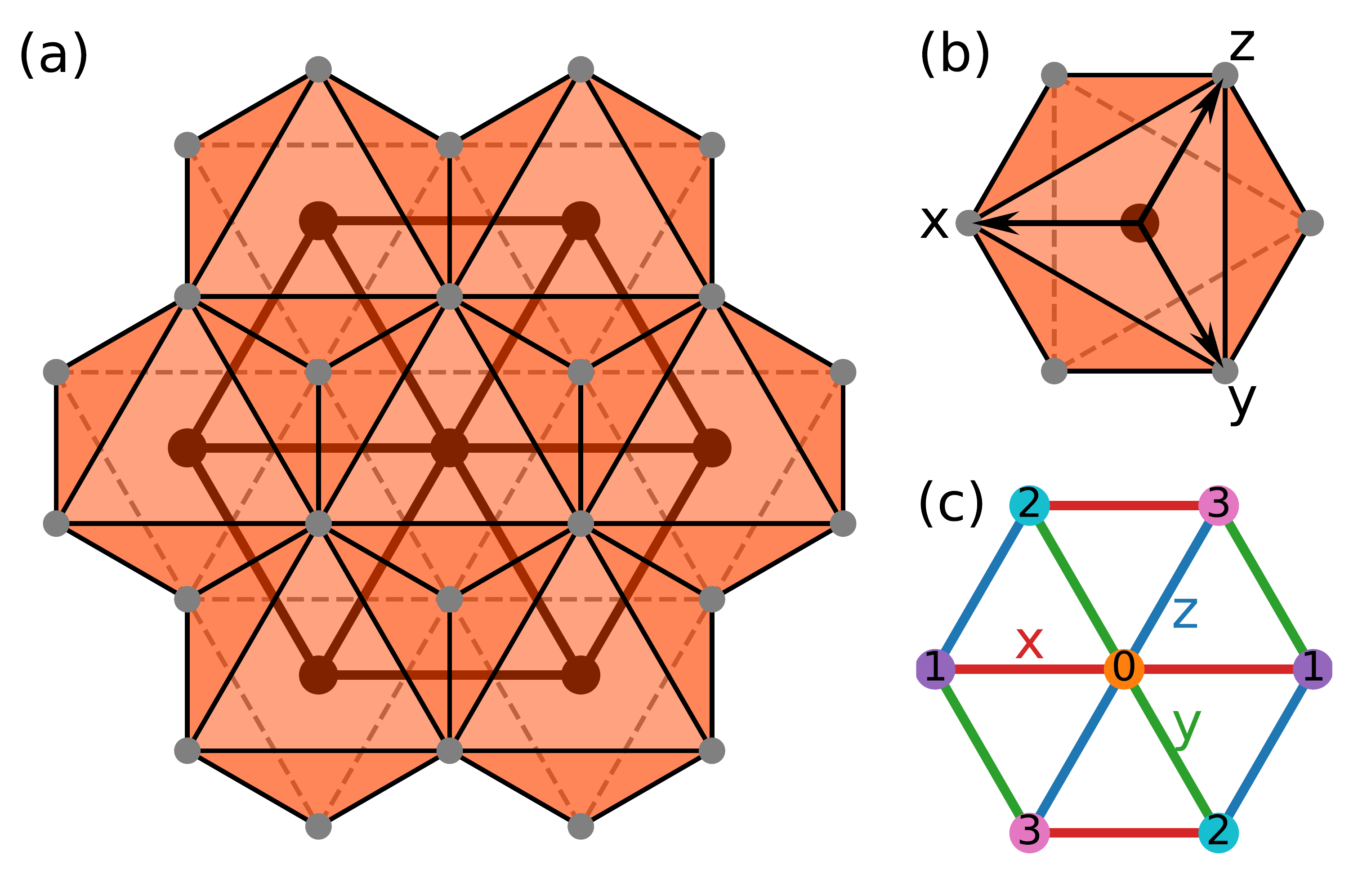}
    \caption{\label{fig:ModelDefinition}(Color online) (a) Top view of the triangular lattice of the edge-sharing octahedron. (b) The orientation of the cubic $x$, $y$, $z$ axes with respect to the octahedron. The spin operators $S^x$, $S^y$ and $S^z$ are defined with respect to this reference frame. (c) Three types of the NN bonds on the triangular lattice, namely $\gamma=x, y, z$ colored red, green and blue, respectively. The different colors of the lattice sites label the four sublattices realizing the four sublattice transformation ( see main text for details).}
\end{figure}
In fact, magnetic ions located at the center of edge-sharing octahedra can not only form the honeycomb lattice but also the triangular lattice (see Fig.~\ref{fig:ModelDefinition}(a)), so the Kitaev and $\Gamma$ terms can naturally be generalized to the triangular lattice~\cite{PhysRevB.89.014414,PhysRevB.93.104417}. On the experimental side, studies on several classes of  compounds containing localized $4d$, $5d$ or $4f$ electrons have recently shown that the quantum spin model on the triangular lattice can be formed by the localized moments~\cite{PhysRevB.86.140405,PhysRevB.94.174410,srep16419,PhysRevLett.115.167203,PhysRevLett.120.087201,CEVALLOS2018154,10.1038/s41535-019-0151-6}. In particular, due to their possible Kramers doublets and spin-orbital coupling, these moments can be treated as $S_{eff}=1/2$ at low temperatures, and the spin-orbital entanglement can induce direction-dependent exchanges, such as the $K$ and $\Gamma$ terms. A typical example is YbMgGaO$_{4}$~\cite{srep16419,PhysRevLett.115.167203}, in which the Yb$^{3+}$ ions form a triangular layer and are surrounded by O$^{2-}$ which construct edge sharing octahedra, but due to the inherent disorder effect there are still controversies about the GS of this material~\cite{srep16419,PhysRevLett.115.167203,PhysRevLett.120.087201,PhysRevB.94.035107,Nature20614,nphys3971,PhysRevLett.117.267202,PhysRevX.8.031028,SciPostPhys.4.1.003,PhysRevB.97.184413}. More recently, an alternative family of compounds AReCh$_2$ (A=alkali, Re=rare-earth, Ch=O, S, Se) with perfect triangular lattices of rare-earth ions have been synthesized and explored. The magnetic susceptibilty and heat capacity data suggest no long-range magnetic order or spin freezing down to the lowest measurement temperature, which implies their candidacy for QSL state~\cite{acsmaterialslett.9b00464,PhysRevMaterials.3.114413,PhysRevB.100.241116,Liu_2018,PhysRevB.99.180401,PhysRevB.100.224417,PhysRevB.100.144432}. These triangluar magnets provide suitable platforms to study the interplay of geometric frustation and exchange frustration induced by spin-orbital couplings.

The phase diagram of the HK model ($\Gamma=0$) on the triangular lattice has been studied by means of Luttinger-Tisza minimization together with classical Monte Carlo simulation~\cite{PhysRevB.93.104417}, exact diagonalization (ED)~\cite{PhysRevB.91.155135,KaiLi2015}, and density matrix renormalization group~\cite{JPSJ.85.114710,PhysRevX.9.021017}. All these methods give consistent results about four magnetically ordered phases: two collinear patterns of FM and stripe types, and two noncollinear patterns of 120$^\circ$ N\'{e}el and noncoplanar spiral types. Note that the distortions of the 120$^\circ$ N\'{e}el order, when the model deviates from the AFM Heisenberg limit ($K=0$, $J>0$), were also called the $Z_{2}$ vortex crystal~\cite{PhysRevB.93.104417,PhysRevB.91.155135,JPSJ.85.114710,PhysRevResearch.1.013002}. However, the nature of the phase around the antiferromagnetic Kitaev point is still under debate. The Luttinger-Tisza minimization method suggested it to be a $Z_{2}$ vortex crystal~\cite{PhysRevB.93.104417}, the Schwinger-fermion mean-field theory proposed it to be a QSL~\cite{KaiLi2015}, the Schwinger-boson mean-field theory thought it to be a nematic phase~\cite{PhysRevB.95.024421}, while the density matrix renormalization group calculations suggested it to be a nematic phase~\cite{PhysRevB.91.155135,JPSJ.85.114710} or a stripe phase~\cite{PhysRevX.9.021017}. When the symmetric off-diagonal $\Gamma$ interaction is included, a classical analysis reveals that the stripe and ferromagnetic phases dominate the $J$-$K$-$\Gamma$ phase diagram, in addition to small regions of 120$^\circ$ N\'{e}el, $Z_{2}$ vortex crystal and nematic phases~\cite{PhysRevB.92.165108}. However, the studies on the effects of quantum fluctuations on the global $J$-$K$-$\Gamma$ phase diagram are scarce. In particular, since no exact solution has been reported so far for the pure spin-$1/2$ Kitaev and $\Gamma$ models on the triangular lattice, it also remains conceptually interesting to investigate whether QSL states could exist as possible GSs due to quantum fluctuations introduced by these exchange-frustrated interactions.

In this paper, we study the global phase diagram and the phase transitions of the triangular lattice $J$-$K$-$\Gamma$ model using a combination of ED, the classical simulation and analytical analyses. In the HK limit ($\Gamma=0$), there are five classical phases: one FM, two stripe, one $120^{\circ}$ N\'{e}el and its dual phases. For the pure AFM Kitaev model, although the system has highly degenerate classical GSs composed of the stripe and nematic states, the order-by-disorder mechanism caused by quantum fluctuations makes the system select the stripe state to be the GS. When the $\Gamma$ term is included, we find five new phases: two FM, one stripe, one modulated stripe and one possible QSL phases. With the aid of classical analysis, we determine the spin configuration for each magnetically ordered phase and explain why there are phase transitions between the phases with the same type of classical orders. On the other hand, we find that the order-by-disorder mechanism makes the GSs of the pure $\Gamma$ models have FM orders, although the classical analyses suggest that the GSs are highly degenerate. For the possible QSL phase, based on the investigation of the spin excitation spectrum, we suggest the GS is a gapped $Z_{2}$ QSL.

\section{Model and Methods}

\subsection{\label{sec:Model}Model}

The $J$-$K$-$\Gamma$ model Hamiltonian on the triangular lattice is given by
\begin{equation}
    H = \sum_{\langle i,j \rangle \in \alpha \beta (\gamma)} \lbrack J \bm{S}_i \cdot \bm{S}_j + K S_i^{\gamma} S_j^{\gamma} + \Gamma (S_i^{\alpha} S_j^{\beta} + S_i^{\beta} S_j^{\alpha}) \rbrack,
    \label{eq:Hamiltonian}
\end{equation}
where $\langle i,j \rangle$ denotes the NN bonds, $\gamma$ takes value $x$, $y$, or $z$ depending on the direction of the NN bond as shown in Fig.~\ref{fig:ModelDefinition}(c), and $\alpha$, $\beta$ are the remaining directions. $J$ and $K$ are the magnitude of the Heisenberg and Kitaev interactions, and $\Gamma$ the symmetric off-diagonal exchanges.

In the followings, for convenience, we fix the energy scale with $\sqrt{J^2 + K^2 + \Gamma^2}=1$ and parametrize the exchange parameters using spherical angles $\alpha$ and $\beta$
\begin{equation}
    J = \sin\alpha \sin\beta, \quad
    K = \sin\alpha \cos\beta, \quad
    \Gamma = \cos\alpha
    \label{eq:Parameters}
\end{equation}
where $\alpha \in [0, \pi]$ and $\beta \in [0, 2\pi]$ to cover the global parameter space.

In the HK limit ($\Gamma=0$), the model~\eqref{eq:Hamiltonian} admits an exact duality transformation, i.e., the so-called four-sublattice-transformation (FST)~\cite{PhysRevB.89.014414}. The FST is a spin rotation transformation which divides the triangluar lattice into four sublattices (see Fig.~\ref{fig:ModelDefinition}(c)) and performs the following rotations of the spins on the four sublattices to map the spin $\bm{S}_{i}$ to $\bm{S}_{i}^{\prime}$,
\begin{align*}
    & \bm{S}_{0}^{\prime} = \bm{S}_{0}& \quad \text{for sublattice 0}, \\
    & \bm{S}_{1}^{\prime} = (S_1^x, -S_1^y, -S_1^z)& \quad \text{for sublattice 1}, \\
    & \bm{S}_{2}^{\prime} = (-S_2^x, S_2^y, -S_2^z)& \quad \text{for sublattice 2}, \\
    & \bm{S}_{3}^{\prime} = (-S_3^x, -S_3^y, S_3^z)& \quad \text{for sublattice 3}.
\end{align*}
This corresponds to a $\pi$ rotation around $x$, $y$, and $z$ axis for the spin operators on the sublattices $1$, $2$, and $3$, respectively. The resulting Hamiltonian $H^{\prime}(\bm{S}^{\prime})$ has the same form as the original Hamiltonian albeit with different model parameters $J^{\prime} = -J$ and $K^{\prime} = 2J + K$. For the spherical angles defined in Eq.~\eqref{eq:Parameters}, the mapping takes the form
\begin{align}
    \tan\beta^{\prime} = -\sin\beta / (2\sin\beta + \cos\beta).
    \label{eq:FST}
\end{align}
This special property of the model can help us to identify some exotic magnetically ordered phases from the well established simple counterparts.

\subsection{\label{subsec:MethodED}Exact Diagonalization Method}

To obtain the quantum phase diagram of the model~\eqref{eq:Hamiltonian}, we perform ED calculations of the GS of the Hamiltonian~\eqref{eq:Hamiltonian} on a $4 \times 6$ cluster with the periodic boundary condition. To detect quantum phase transitions, the second derivatives of the GS energy, $-\partial^2E_0/\partial\alpha^2$ and $-\partial^2E_0/\partial\beta^2$ were computed and its singularities are used to identify possible phase transtions.

To identify the ground-state properties, we first examine the static structure factor (SSF),
\begin{align}
    \mathcal{S}(\bm{Q}) = \frac{1}{N} \sum_{ij} \langle \Omega | \bm{S}_i \cdot \bm{S}_j | \Omega \rangle e^{i \bm{Q} \cdot (\bm{R}_i - \bm{R}_j)},
    \label{eq:SSF}
\end{align}
from which we can find the wave vectors of the ordered phases and distinguish possible QSL states. Here, $|\Omega\rangle$ is the GS, $N$ is the total number of lattice sites, and $\bm{R}_i$ the position of site $i$.

To further determine the magnetic configurations of the magnetically ordered phases, we employ a method by studying the projections of the exact GSs of the finite cluster to the classical states~\cite{PhysRevB.94.064435}. The basic idea of this method is to measure the probabilities of the cluster spin coherent states in the exact cluster GS with varying moment directions. The cluster spin coherent state is a direct product of spin-1/2 coherent states on each site $i$, i.e.,
\begin{align}
    | \Psi \rangle = \bigotimes_{i=1}^{N} | \theta_{i}, \phi_{i} \rangle,
    \label{eq:ClusterCoherentState}
\end{align}
where the spin-1/2 coherent state
\begin{align}
    | \theta, \phi \rangle = \mathcal{R}_{z}(\phi) \mathcal{R}_{y}(\theta) | \uparrow \rangle = e^{-i \phi S^z} e^{-i \theta S^y} | \uparrow \rangle
    \label{eq:Spin-1/2CoherentState}
\end{align}
is fully polarized state along the $(\theta, \phi)$ direction. Here the cubic axes are used (see Fig.~\ref{fig:ModelDefinition}(b)), $\theta$ and $\phi$ are the conventional spherical angles. By calculating the overlap between the exact cluster GS and cluster spin conherent states, $P = | \langle \Psi | GS \rangle |^2$, and maximizing its value with respect to $\theta$'s and $\phi$'s, we can then identify the classical spin pattern that best fits the exact quantum GS.

Since one of the key characteristics of QSL is the fractional excitation, which can lead to a continuous spectrum, so we study the dynamic structure factor (DSF) $A(\bm{k}, \omega)$ to search for the possible QSLs. $A(\bm{k}, \omega)$ is given by
\begin{align}
    & A(\bm{k}, \omega) = -\frac{1}{\pi} \text{Im} \mathcal{S}(\bm{k}, \omega), \label{eq:Akomega} \\
    & \mathcal{S}(\bm{k}, \omega) = \frac{1}{N} \sum_{ij} \mathcal{S}_{ij}(\omega) e^{i \bm{k} \cdot (\bm{R}_i - \bm{R}_j)}, \\
    & \mathcal{S}_{ij}(\omega) = \langle \Omega | S_i^{+} \frac{1}{\omega + i0^{+} - H + E_0} S_j^{-} | \Omega \rangle, \label{eq:GreenFunction}
\end{align}
where $E_0$ is the GS energy.

\subsection{\label{subsec:MethodMC}Classical Monte Carlo Method}

In order to better understand the quantum phases identified by the ED method, we also perform classical Monte Carlo simulations. We begin with paralleling-tempering Monte Carlo~\cite{Hukushima96} on 40 replicas with temperature $T/J$ ranging from $0.001$ to $1.0$. For each replica, we sample it with a combination of heat-bath~\cite{Miyatake84} and over-relaxation method~\cite{Berg} mainly on a $24 \times 24$ triangle lattice with periodic boundary conditions. A whole Monte Carlo step consists of a single heat-bath sweep and subsequent 10 over-relaxation sweeps over the entire lattice. We perform $10^6$ Monte Carlo steps per replica, then, we copy out the spin configuration from the lowest-T replica, and sample it with a combination of zero-temperature heat-bath and over-relaxation method to get the GS. The zero-temperature heat-bath sampling is simply aligning the spins according to their local fields:
\begin{align}
    \bm{S}_{i} = \frac{\bm{h}_{i}^{loc}}{|\bm{h}_{i}^{loc}|} S,
\end{align}
with
\begin{align}
    \bm{h}_{i}^{loc} = \sum_{\langle j \rangle} J \bm{S}_{j} + K {S}^{\gamma}_{j} \hat{e}^{\gamma} + \Gamma (S^{\alpha}_{j} \hat{e}^{\beta} + S^{\beta}_{j} \hat{e}^{\alpha}),
\end{align}
For some competing states, we start from several different initial configurations in order to obtain the correct classical GS.

From the magnetic configuration of GS, we compute the SSF given by
\begin{align}
    \mathcal{S}_{\bm{k}} = \frac{1}{N} \sum_{ij} \left \langle \bm{S}_i \cdot \bm{S}_j \right \rangle e^{i \bm{k} \cdot (\bm{r}_i-\bm{r}_j)},
\end{align}
which is a key characteristic to identify magnetic phases.

\section{\label{sec:Results}Results}

\subsection{\label{subsec:GlobalPhaseDiagram}Global Phase Diagram}

\begin{figure*}
    \centering
    \includegraphics[width=0.95\textwidth]{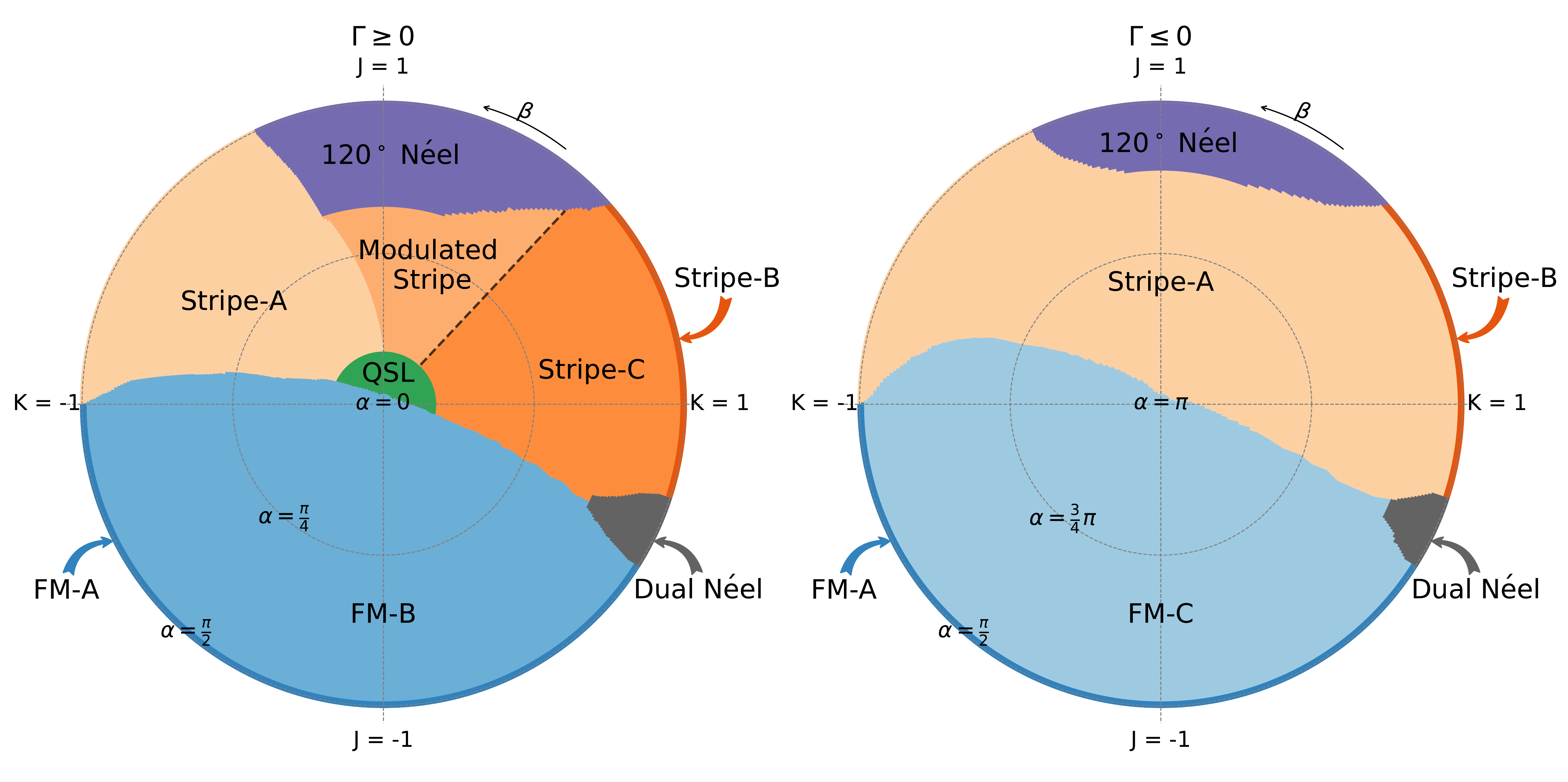}
    \caption{\label{fig:QuantumPhaseDiagram}(Color online) Global phase diagram of the triangular lattice $J$-$K$-$\Gamma$ model. The angle $\alpha$ and $\beta$ denote the radial and azimuthal angles, respectively. There are in total ten phases including three FM phases denoted as FM-A, FM-B and FM-C, three stripe phases denoted as Stripe-A, Stripe-B and Stripe-C, a modulated stripe phase, a 120$^\circ$ N\'{e}el phase, a dual N\'{e}el phase and a possible QSL phase. Phase boundaries are determined by the singularities of $-\partial^2E_0/\partial\alpha^2$ and $-\partial^2E_0/\partial\beta^2$ from ED calculations except for that depicted by the dashed line between the modulated stripe and stripe-C phases given by the classical analyses.}
\end{figure*}

The quantum phase diagrams obtained by the ED method for $\Gamma \geq 0$ and $\Gamma \leq 0$ on a $4 \times 6$ cluster with the periodic boundary condition are presented in Fig.~\ref{fig:QuantumPhaseDiagram}, where the phase boundaries are determined by the location of singularities in $-\partial^2E_0/\partial\alpha^2$ and $-\partial^2E_0/\partial\beta^2$ (see Appendix \ref{apx:PhaseBoundary} for details). We find nine magnetically ordered phases and a phase ascribed to be a QSL. The nine ordered magnetic phases consist of three stripe phases (dubbed as stripe-A, strip-B and stripe-C in Fig.~\ref{fig:QuantumPhaseDiagram}), three ferromagnetic phases (dubbed as FM-A, FM-B and FM-C in Fig.~\ref{fig:QuantumPhaseDiagram}), one 120$^\circ$ N\'{e}el phase, one dual N\'{e}el phase and one modulated stripe phase. In the followings, we will discuss the details on the natures of these phases and the corresponding phase transitions.

\begin{figure}
    \centering
    \includegraphics[width=\columnwidth]{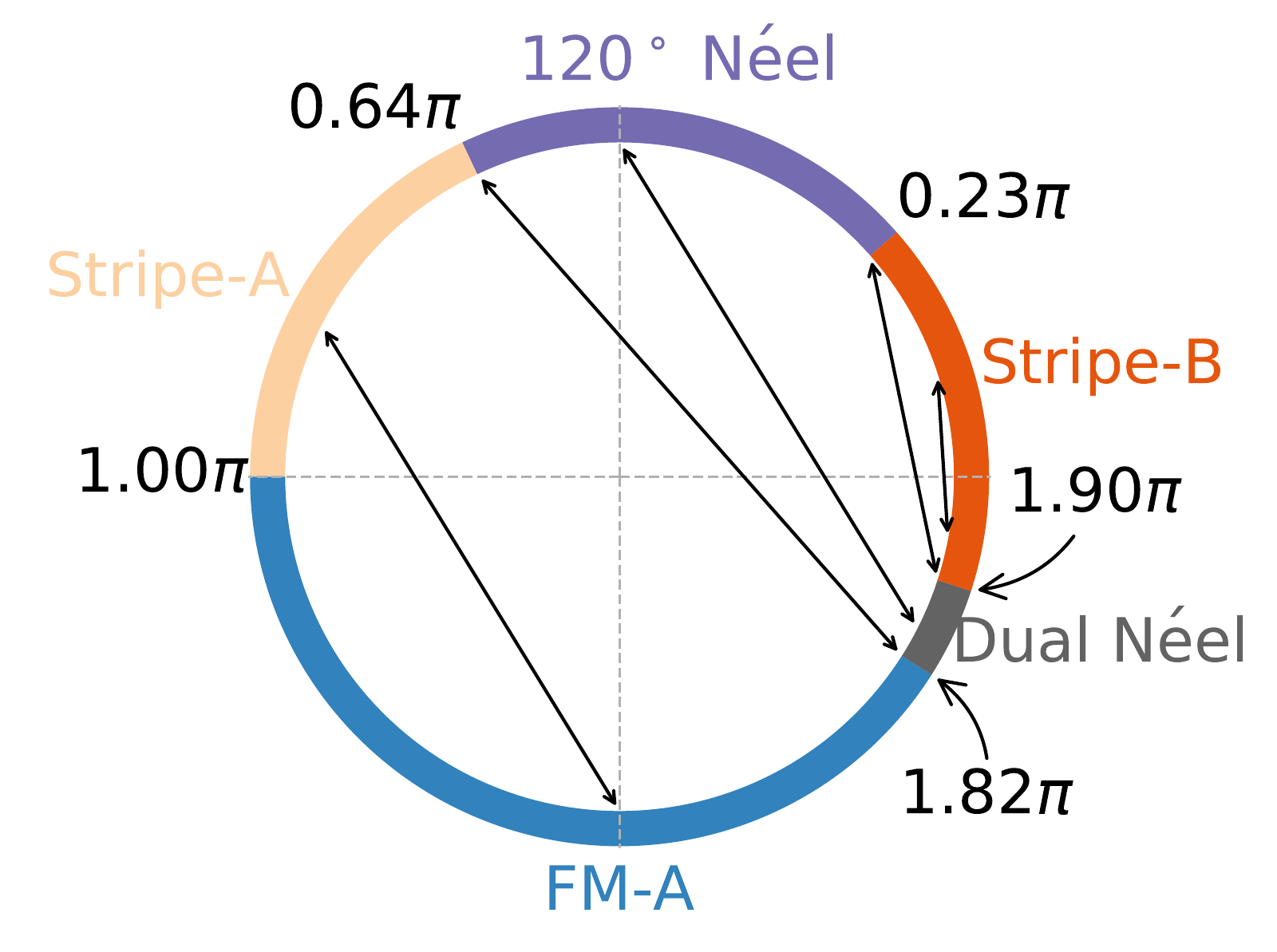}
    \caption{\label{fig:HKModel}(Color online) Phase diagram in the HK limit ($\Gamma=0$). The double-arrow lines are the representative lines that connect the points exactly related by the FST.}
\end{figure}
Let's first discuss the HK limit for $\Gamma=0$. In accord with the previous study~\cite{KaiLi2015}, we find five quantum phases (see Fig.~\ref{fig:HKModel}). For pure Heisenberg models ($K=\Gamma=0$), the GSs are well known, i.e., a FM state (denoted as FM-A) for $J<0$ and a 120$^\circ$ N\'{e}el state for $J>0$, respectively. As noted in Sec.~\ref{sec:Model}, the HK model ($\Gamma=0$) preserves its form under the FST~\cite{PhysRevB.89.014414} but with different exchange interactions. Thus, by virtue of the FST, we can identify other two magnetically ordered phases. For the FM Heisenberg model of the rotated spin operators $\bm{S}^{\prime}$ whose GS is a FM state, the FST maps it to a Heisenberg-Kitaev model of the original spin operators $\bm{S}$ with $\beta=\pi - \arctan\frac{1}{2}$. Accordingly, the FM state of $\bm{S}^{\prime}$ is transformed to a collinear stripe state (denoted as stripe-A) of $\bm{S}$. Similarly, the AFM Heisenberg model of $\bm{S}^{\prime}$ is also mapped to a Heisenberg-Kitaev model of $\bm{S}$ with $\beta=-\arctan\frac{1}{2}$, and the 120$^\circ$ N\'{e}el state of $\bm{S}^{\prime}$ is transformed to a noncollinear spiral order (Dual N\'{e}el) of $\bm{S}$. Moreover, as indicated in Fig.~\ref{fig:HKModel}, the transition points $\beta=0.23\pi$ and $0.64\pi$ can also be well mapped to the transition points $\beta=1.90\pi$ and $1.82\pi$ through the FST. The isolated transition point $\beta=\pi$ is also consistent with the FST, as it is mapped to itself under the FST.

\begin{figure}
    \centering
    \includegraphics[width=\columnwidth]{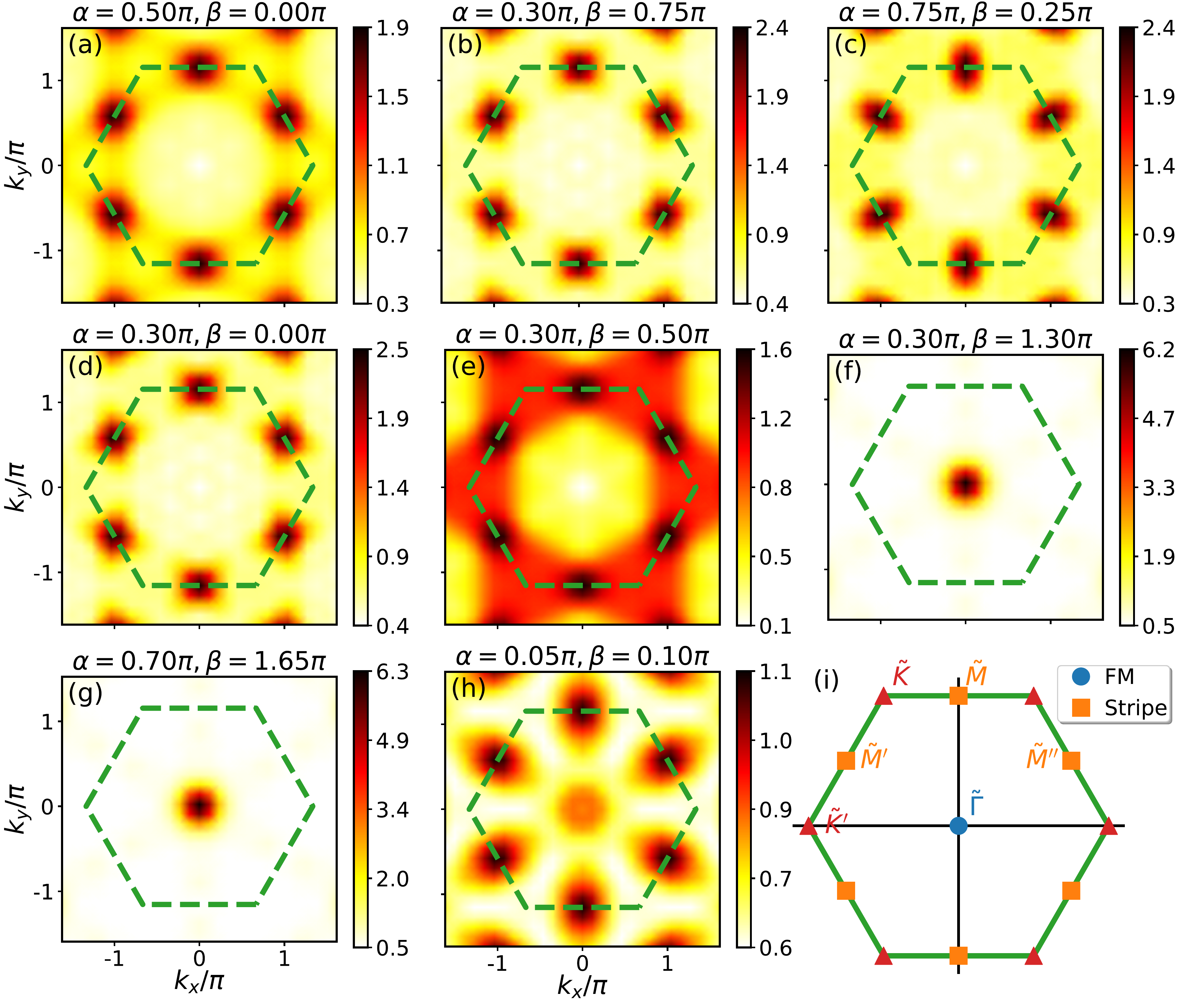}
    \caption{\label{fig:StructureFactors}(Color online) SSFs from the ED calculations for representative interaction parameters in different phases: (a) strip-B, (b) and (c) stripe-A, (d) stripe-C, (e) modulated stripe, (f) FM-B, (g) FM-C, (h) QSL. The green dashed lines marks the first BZ of the triangular lattice. (i) Characteristic wave vectors for the FM and stipe phases. See Appendix~\ref{apx:SSF} for the SSFs of the 120$^\circ$ N\'{e}el phase, dual N\'{e}el phase and pure $\Gamma$ models.}
\end{figure}

For the phase near the AFM Kitaev point ($J=\Gamma=0$, $K=1$), which is between $\beta=1.90\pi$ and $0.23\pi$,  the points in this phase are still mapped to those in the same phase under the FST, and it was supposed to be a magnetically disordered phase~\cite{KaiLi2015}. However, as shown in Fig.~\ref{fig:StructureFactors}(a), when we check the SSF in this phase, it is found that the SSF shows obvious peaks at the $\tilde{M}$ points, which implies that it is likely to be a magnetically ordered phase. We further perform a ``basin-hopping" global optimization~\cite{jp970984n} on a $12 \times 12$ lattice by taking the spins as classical magnetic moments to search for the possible spin configurations under this set of interaction parameters. We find that there are two types of degenerate classical GSs, one of which has the stripe order and the other the nematic order, since the AFM chains are decoupled (see Appendix~\ref{apx:DegeneratedStates} for details). When we consider the quantum fluctuations generated by the Kitaev interactions not along the AFM chains, the $4^{th}$ order corrections will give an effective coupling between the next-nearest-neighbor spins~\cite{PhysRevB.92.184416}, which can stabilize the stripe order. As shown in Fig.~\ref{fig:StructureFactors}(i), the peaks of SSF of the classical stripe orders are also consistent with the result calculated from ED [see Fig.~\ref{fig:StructureFactors}(a)]. Thus, the degeneracy of the spin configurations is lifted by the order-by-disorder mechanism, and we can identify that the phase containing the AFM Kitaev point of the HK model is a stripe phase, which we label the stripe-B. Moreover, every stripe order in this phase (see Appendix~\ref{apx:DegeneratedStates} for details) can be transformed to another degenerate stripe pattern in the same phase by a FST, so these stripe orders are consistent with the FST.

Then, let us study the effects of the $\Gamma$ term on the phase diagram. One may expect that the phases in the HK limit ($\Gamma=0$) would extend to a finite region in the global phase diagram of the $J$-$K$-$\Gamma$ model. Our ED results do show that the stripe-A phase can extend to a large region from $\Gamma>0$ to $\Gamma<0$, in particular it extends to the region connecting $K=-1$ to $K=1$ for $\Gamma<0$. The region of the dual N\'{e}el phase that was dubbed as a dual-$Z_{2}$ vortex crystal phase in Ref.~\onlinecite{PhysRevB.92.165108} also survives
the introduction of the $\Gamma$ term, though the region is much smaller than the stripe-A phase. These results
are consistent qualitatively with the classical results~\cite{PhysRevB.92.165108}. However, we find that an infinitesimal $\Gamma$ interaction can make the FM-A phase and the stripe-B phase unstable, so these two phases are actually phase boundaries in the global phase diagram. On the other hand, although the 120$^\circ$ N\'{e}el phase can also extend to a certain region, its area in the phase diagram is much smaller than the classical results~\cite{PhysRevB.92.165108}. Besides these phases that already exist in the HK limit, there are five other new phases according to our ED calculations. Since the SSF is a key physical quantity to reveal the nature of each quantum phases, especially the spin configuration of the ordered phases, we will discuss the properties of the phases with $\Gamma\neq0$ in detail according to the SSFs.

In Fig.~\ref{fig:StructureFactors}(b) and (c), we show the SSFs for two typical points in the stripe-A phase, i.e. $\alpha=0.3\pi$, $\beta=0.75\pi$ and $\alpha=0.75\pi$, $\beta=0.25\pi$ in the $\Gamma>0$ and $\Gamma<0$ regions, respectively. Both of the SSFs show obvious peaks at the $\tilde{M}$ points, which is a typical characteristic of the stripe order as shown in Fig.~\ref{fig:StructureFactors}(i). Thus, this further confirms that the stripe-A phase extents to a large region in the phase diagram. In Fig.~\ref{fig:StructureFactors}(d), we illustrate the SSF for $\alpha=0.3\pi$, $\beta=0$ for one of the new phase that does not exist in the HK limit, which also shows distinct peaks at the $\tilde{M}$ points, so the phase containing this point is also a stripe phase and we denote it the stripe-C phase in the phase diagram.

\begin{figure}
    \centering
    \includegraphics[width=0.95\columnwidth]{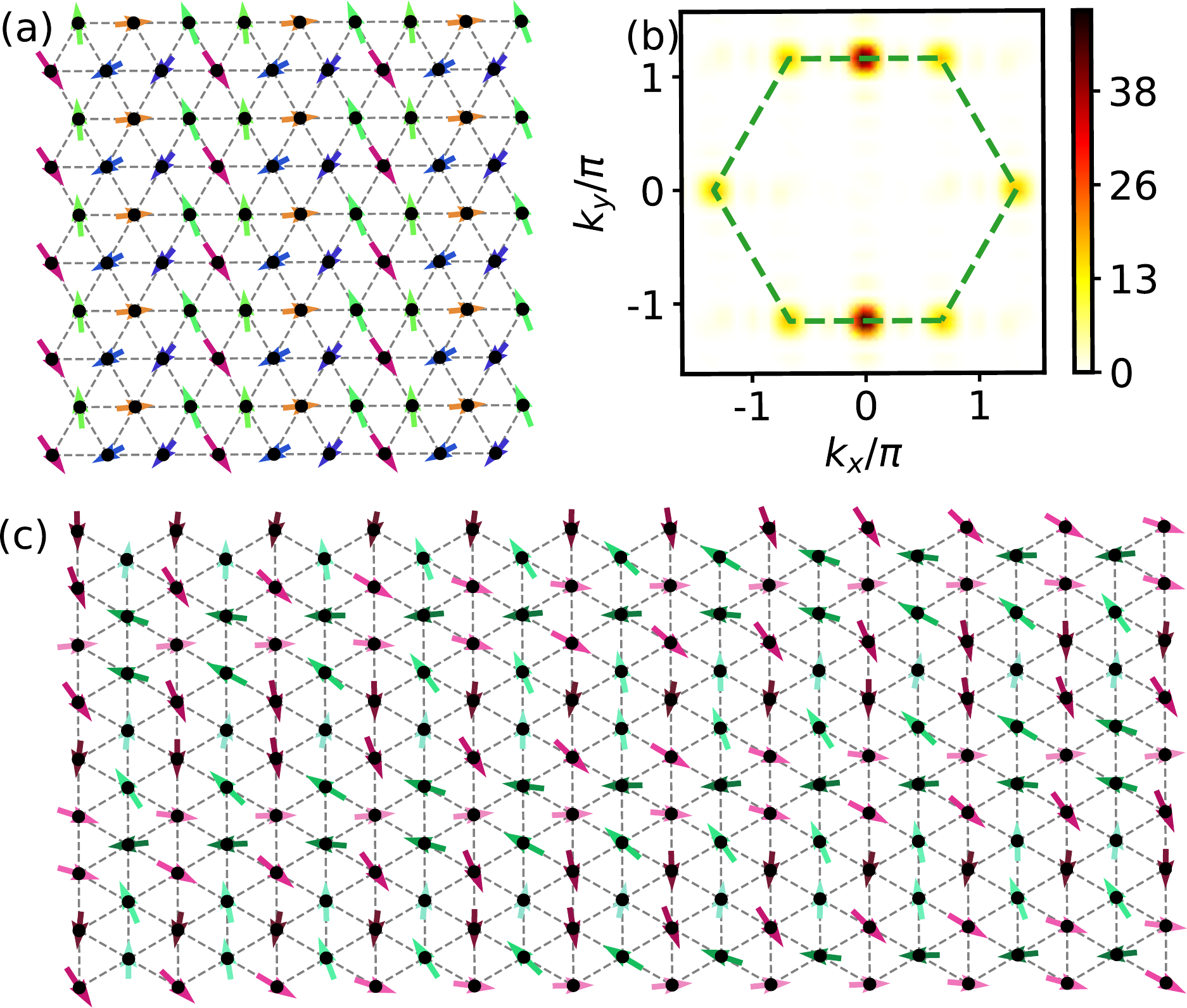}
    \caption{\label{fig:ModulatedStripe}(Color online) (a) Spin configuration obtained from basin-hopping optimization for $\alpha=0.3\pi$, $\beta=0.5\pi$ in a $12 \times 12$ cluster. (b) SSF corresponding to the spin configuration shown in (a). (c) Spin configuration obtained from the classical Monte Carlo calculations in a $24 \times 24$ cluster.}
\end{figure}

When we carefully examine the SSF near $\beta=0.5\pi$, we find that the SSF is obviously different from those of the stripe phases. As shown in Fig.~\ref{fig:StructureFactors}(e) for $\alpha=0.3\pi$ and $\beta=0.5\pi$, besides the peaks at the $\tilde{M}$ points, the SSF also shows significant intensities along the boundary of the BZ. In order to clarify whether there is another magnetic ordering phase near $\beta=0.5\pi$, we use the ``basin-hopping" technique again to search for the possible spin order for $\alpha=0.3\pi$ and $\beta=0.5\pi$ on a $12 \times 12$ lattice. The spin configuration is depicted in Fig.~\ref{fig:ModulatedStripe}(a), which exhibits a modulated stripe order. For this modulated stripe order, the spins are ferromagnetic along a next-nearest-neighbor direction and the spin orientation of the ferromagnetic chains are modulated to form a structure with a period of six. As illustrated in Fig.~\ref{fig:ModulatedStripe}(b), the SSF of this classical six-chain stripe order shows clear peaks at the $\tilde{K}$ points in addition to those at the $\tilde{M}$ points, which is consistent with the characteristic of the SSF in Fig.~\ref{fig:StructureFactors}(e) calculated by the ED. We further perform a classical Monte Carlo simulations (see Sec.~\ref{subsec:MethodMC}) on a $24 \times24$ lattice to check the stability of the modulated stripe order in the large-size systems, and the ground-state spin configuration is shown in Fig.~\ref{fig:ModulatedStripe}(c). We find that the spin arrangement is still a modulated stripe order, but the spin orientations are also modulated along the ferromagnetic chains in addition to a larger modulated period perpendicular to the chains. Thus, we divide a transition region between the stipe-A and stripe-C phase and term it the modulated stripe order phase.

For the other two phases with large areas and mainly located in the $J<0$ region of the phase diagram, we show the SSFs of two representative points with $\alpha=0.3\pi, \beta=1.3\pi$ and $\alpha=0.7\pi, \beta=1.65\pi$ in Fig.~\ref{fig:StructureFactors}(f) and (g), respectively. Since the SSFs are peaked at $\tilde{\Gamma}$ point of the first BZ, so both phases have ferromagnetic orders and they are denoted with FM-B and FM-C in Fig.~\ref{fig:QuantumPhaseDiagram}. We note that there are two special points with $\Gamma=\pm1$ ($J=K=0$) in the two FM phases, which are located at the phase boundary in the classical phase diagram~\cite{PhysRevB.92.165108}. Thus, it is worthy to discuss the properties of the pure $\Gamma$ models in details, and we perform the classical analyses by using the ``basin-hopping" global optimization and classical Monte Carlo methods. For $\Gamma=+1$, apart from the FM GS with the spin orientation lying in the lattice plane, we also found disordered states having the same energy as the FM state (see Appendix~\ref{apx:DegeneratedStates} for details). For $\Gamma=-1$, in addition to the FM order perpendicular to the lattice plane, there are also several noncollinear magnetic orders energetically degenerate with the FM state (see Appendix~\ref{apx:DegeneratedStates} for details). However, according to our ED calculations, for the quantum model, quantum flucations select the FM ordered states as the GS out of the degenerate manifolds of classical states. This selection of states among the degenerate classical ground states is the so called order-by-disorder mechanism, which has previously been applied in a number of insulating magnets, especially in cases where frustration leads to a degenerate manifold of classical ground state configurations that is broken by quantum fluctuations~\cite{JPSJ.54.4494,PhysRevLett.62.2056,PhysRevLett.88.067203,PhysRevB.81.214419,PhysRevLett.109.077204}.

According to the above analyses, we find that, unlike the case for the honeycomb lattice, the Kitaev interactions on the triangular lattice do not give rise to QSL states. Here, the large coordination number plays a key role to stabilize the classical magnetic orders, although the geometric and exchange frustrations coexist in the triangular lattice $J$-$K$-$\Gamma$ model. However, the SSF for the quantum phase in a small region around the point with $J=K=0$ and $\Gamma=1$ illustrated in Fig.~\ref{fig:StructureFactors}(h) exhibits high intensities at both $\tilde{\Gamma}$ and $\tilde{M}$ points, and it seems impossible for a classical magnetically ordering state to satisfy these two wave vectors simultaneously, so we infer that this quantum phase is a QSL candidate.

So far, we have basically determine the nature of each phase, but there are still two problems should be explained: first, why there are phase transitions between the phases with the same type of classical orders, such as the phase transitions between FM phases or stripe phases, and why the FM and stripe phases of the HK model are the phase boundaries in the global phase diagram of $J$-$K$-$\Gamma$ model; second, what other important characteristics of the possible QSL phase in the phase diagram can be used to help us understand its properties more deeply. In the followings, we will address these problems.

\subsection{\label{subsec:FMPhases}FM Phases}

To have a better understanding of the phase transitions between these FM phases, we first study the classical $J$-$K$-$\Gamma$ model where the spin operators are viewed as unit-vectors in the three dimension. For classical FM states, all spins are aligned in parallel and the energy per lattice site is given by
\begin{equation}
    E_{FM}^{c} = (3J + K) + 2 \Gamma (v^x v^y + v^y v^z + v^z v^x) \label{eq:EcFM}
\end{equation}
where $v^x$, $v^y$, $v^z$ are the three components of the classical moment vector. On this level, the moment direction of the FM state is determined solely by $\Gamma$, so the problem becomes finding the global minimum and maximum of the multi-variable function $f(v^x, v^y, v^z) = v^x v^y + v^y v^z + v^z v^x$ with the constraint $|\bm{v}| = 1$. $f(v^x, v^y, v^z)$ takes the maximum value $f_{max}=1$ at $v^x=v^y=v^z=\pm 1/\sqrt{3}$ and minimum value $f_{min}=-0.5$ when the conditions $v^x + v^y + v^z = 0$ and $|\bm{v}| = 1$ are fulfilled. The condition $v^x + v^y + v^z = 0$ specifies a plane perpendicular to the $[111]$ direction, and considering the reference frame shown in Fig.~\ref{fig:ModelDefinition}(b) it is actually the plane of the triangular lattice. That is to say, for $\Gamma > 0$ the ordered moment of the classical FM state prefers to lie in the lattice plane, while for $\Gamma<0$ the ordered moment would perpendicular to the lattice plane. This is the reason that there is a phase transition between the FM-B and FM-C phases, since the dependence of the ground state energy on the interaction parameters is different for the two FM phases, and the phase boundary is the line with $\Gamma=0$.

\begin{figure}
    \centering
    \includegraphics[width=\columnwidth]{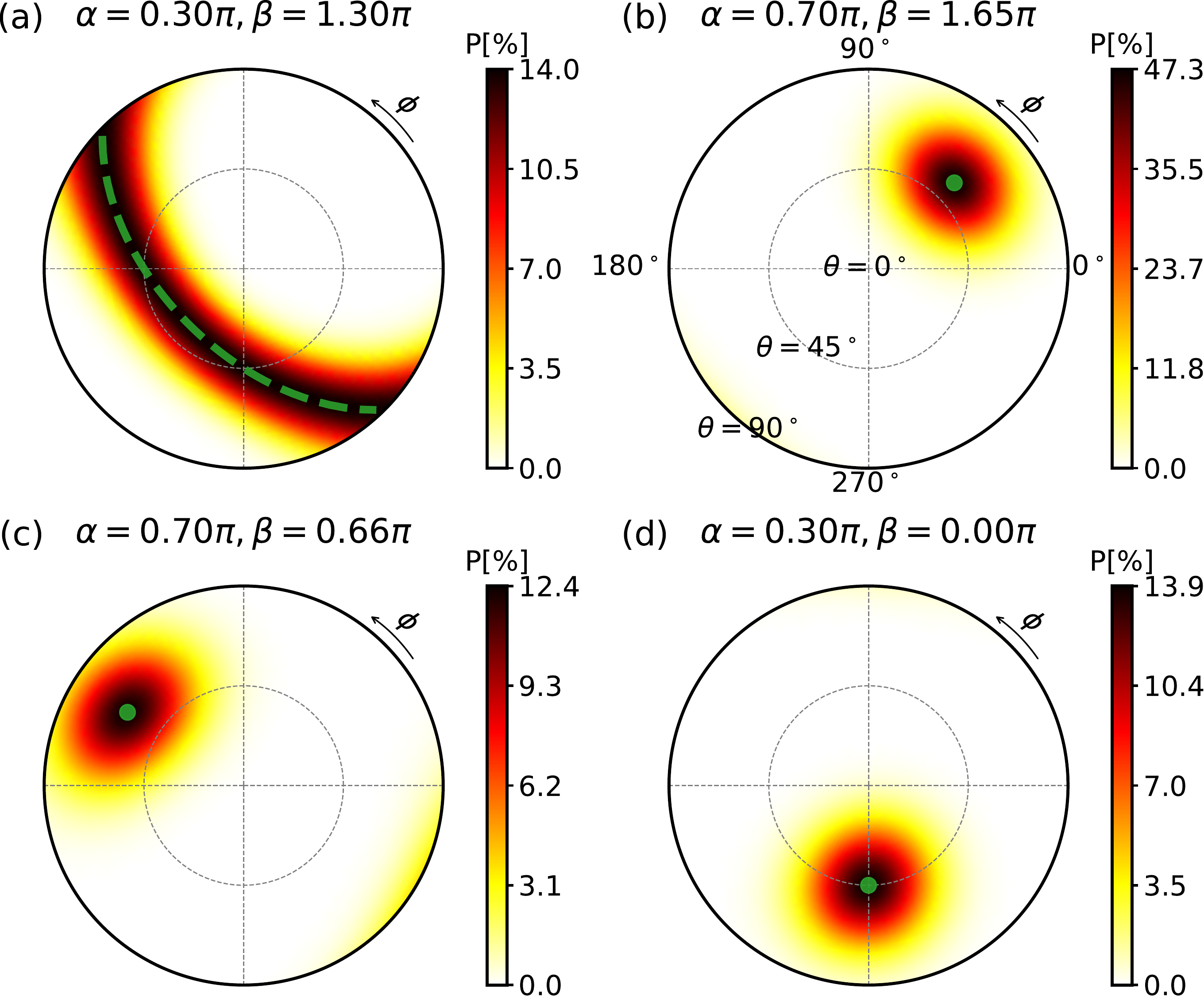}
    \caption{\label{fig:Proabilities}(Color online) Maps of the probabilities of the cluster spin coherent states given by Eq.~\eqref{eq:Spin-1/2CoherentState} in exact cluster GS for (a) FM-B, (b) FM-C, (c) stripe-A and (d) stripe-C phases. The radial and polar coordinate gives the angles $\theta$ and $\phi$, which are spherical angles with respect to the reference frame shown in Fig.~\ref{fig:ModelDefinition}(b). The green dashed lines and solid circles mark the ordered moment directions for the classical states. Note that the GSs are two-fold degenerate for the FM-C phase, and six-fold degenerate for the stripe-A and stripe-C phases, so the total probabilities for all of these phases are approximate to $1$. }
\end{figure}

In order to confirm the consistency between the above classical analyses and the ED results, we use the spin coherent state to extract the moment direction of these FM phases from our ED cluster GS~\cite{PhysRevB.94.064435}. Since the cluster spin coherent state defined in Eq.~(\ref{eq:Spin-1/2CoherentState}) is captured only by a single pair of $(\theta, \phi)$ for the collinear states, it is easy to determine the direction of the FM orders by inspecting the probability map $P(\theta, \phi) = | \langle \Psi (\theta, \phi) | GS \rangle |^2$. As shown in Fig.~\ref{fig:Proabilities}(a), for the FM-B phase, the peaks of the probability form a ring, which indicates that the moment being constrained to a plane with all directions degenerate. To compare with the classical results, we plot the classical magnetic moment directions with the green dashed line in Fig.~\ref{fig:Proabilities}(a), and we find that the the probability is concentrated near the classical magnetic moment directions. For the FM-C phase, the probability map is clearly peaked at specific directions, which are also consistent with the classical results marked by the green solid circle in Fig. \ref{fig:Proabilities}(b). Moreover, the large overlaps between the exact cluster GSs and FM cluster spin coherent states again provide solid evidences that the ED results are consistent with the classical analyses.

\subsection{\label{subsec:StripePhases}Stripe Phases}

Inspired by the previous discussion of FM phases, we here perform the same analysis for the stripe phases. For stripe order, there are three degenerated spin configurations in real-space, designated as stripe-$x$, stripe-$y$ and stripe-$z$, where the FM chains are along the $x$-bond, $y$-bond and $z$-bond directions, respectively. For simplicity, we take the stripe-$x$ configuration as an example, and the other two configurations can be obtained by analogy. The energy per lattice site is
\begin{eqnarray}
    E_{stripe-x}^{c} & = & -(J + K) + 2 K v^x v^x \nonumber \\
        & & +\: 2 \Gamma (v^y v^z - v^z v^x - v^x v^y).
        \label{eq:EcStripeX}
\end{eqnarray}
The ordered moment direction of the classical stripe order is determined by the interaction parameters $K$ and $\Gamma$. In general ($K,\Gamma \neq 0$), $E_{stripe-x}^{c}$ has six extreme points where the first derivatives with respect to $v_x$, $v_y$ and $v_z$ are equal to zero. Here we give three of them explicitly and the other three are opposite to the given ones,
\begin{subequations}
    \label{eq:whole}
    \begin{eqnarray}
        & \bm{v}_0:& \quad v_{0}^{y}=-v_{0}^{z} = 1/\sqrt{2}, \quad v_{0}^{x} = 0, \label{eq:v0} \\
        & \bm{v}_1:& \quad v_{1}^{y}=v_{1}^{z} = f_{1}(K, \Gamma), \quad v_{1}^{x} = g_{1}(K, \Gamma) v_{1}^{y}, \label{eq:v1} \\
        & \bm{v}_2:& \quad v_{2}^{y}=v_{2}^{z} = f_{2}(K, \Gamma), \quad v_{2}^{x} = g_{2}(K, \Gamma) v_{2}^{y}, \label{eq:v2}
    \end{eqnarray}
\end{subequations}
where
\begin{align}
& f_{1,2}(K, \Gamma)= \frac{|\Gamma|}{\sqrt{4 \lambda_{1,2} (\lambda_{1,2} + \Gamma) + 3 \Gamma^{2}}}, \nonumber \\
& g_{1,2}(K, \Gamma)= (2 \lambda_{1,2} + \Gamma) / \Gamma, \nonumber
\end{align}
with
\begin{align}
& \lambda_{1}=- (\Gamma +2 K - \sqrt{9\Gamma^2 - 4 \Gamma K + 4 K^2}) / 4, \nonumber \\
& \lambda_{2}=-(\Gamma +2 K + \sqrt{9\Gamma^2 + 4 \Gamma K + 4 K^2}) / 4. \nonumber
\end{align}

In the HK limit, for $K>0$, it can be clearly seen from Eq.~\eqref{eq:EcStripeX} that $E_{stripe-x}^{c}$ takes minimum value at $v^x = 0$ which means that stripe-$x$ prefers to lie in the $yz$ plane. An infinitesimal positive $\Gamma$ would fix the ordered moment to the $\bm{v}_0$ direction, whereas the negative $\Gamma$ drives it to the $\bm{v}_1$ direction. Thus, there is a phase transition between the stripe-A and stripe-C phases, as the ground state energies have different dependence on the interaction parameters for the two stripe phases, and the phase boundary is the line with $\Gamma=0$.

To confirm that the classical analysis is consistent with our ED calculations, we use the spin coherent states again to extract the ordered moment direction of the ED ground states in these stripe phases. For brevity, we here only construct cluster spin coherent state based on the stripe-$x$ and present the resulting probability maps in Fig.~\ref{fig:Proabilities}(c) and (d) for stripe-A and stripe-C, respectively. We can see that there are large overlaps between the exact cluster GS and stripe cluster spin coherent state, which indicates that the results based on the ED calculation is consistent with the classical analysis.

\subsection{\label{subsec:QSL}Possible QSL}

\begin{figure}
    \centering
    \includegraphics[width=\columnwidth]{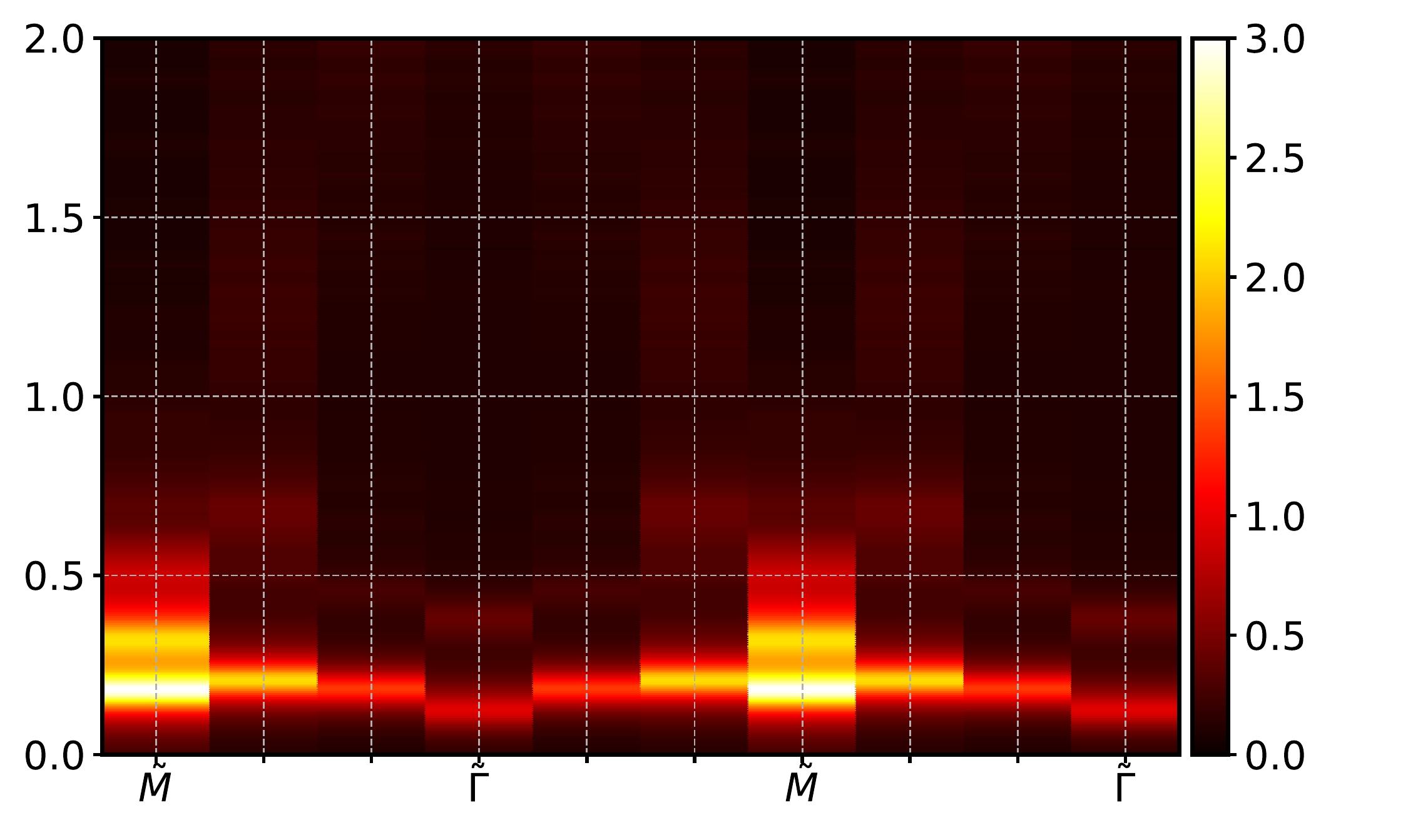}
    \caption{\label{fig:Spectrum}(Color online) DSF $A(\bm{k}, \omega)$ for $\alpha=0.05\pi,\beta=0.1\pi$ in the QSL phase. The path $\tilde{M}-\tilde{\Gamma}-\tilde{M}-\tilde{\Gamma}$ is shown in Fig.~\ref{fig:StructureFactors}(i).}
\end{figure}

Although the GSs of the pure Heisenberg, Kitaev and $\Gamma$ models are not QSL, the interplay of the Heisenberg, Kitaev and $\Gamma$ interactions may induce a possible QSL phase near the $\Gamma=1$ limit as discussed above according to the SSF. To have further understanding of the nature of this quantum phase that does not exist in the classical phase diagram~\cite{PhysRevB.92.165108}, we also calculate the DSF $A(\mathbf{k}, \omega)$, which in general can provide some critical information about the properties of a QSL. A typical $A(\mathbf{k}, \omega)$ profile is shown in Fig~\ref{fig:Spectrum}. We can find that a significant feature is that the whole spectrum is a broad continuum, which is a characteristic of QSL originating from the fractionalization of the $S=1$ spin excitations. Thus, the DSF gives another evidence that this quantum phase is a QSL. The DSF also shows another feature that the periodicity of its lower edge is doubled (i.e., $\tilde{M}$-$\tilde{\Gamma}$ forms a period instead of the usual $\tilde{M}$-$\tilde{M}$), which is a signature of the translational symmetry fractionalization of a $Z_{2}$ spin liquid~\cite{PhysRevB.90.121102,mei2015fractionalized,PhysRevB.99.205119}. Moreover, we also find that there is an obvious gap in the DSF. Thus, combined with all of the above results, we propose the GS of quantum $J$-$K$-$\Gamma$ model at the green area of the phase digram in Fig.~\ref{fig:QuantumPhaseDiagram} is a gapped $Z_{2}$ QSL.

\section{\label{sec:Summary}Summary}

%In summary, by using exact diagnoalization calculations and classical analyses, we map out the global phase diagram of the $J$-$K$-$\Gamma$ model. The global phase diagram of the quantum model is analogus to its classical counterpart except a small area of QSL near the positive $\Gamma$ point, for which we propose the ground state to be a gapped $Z_{2}$ QSL based on its spin excitation spectrum. We also provide detailed descriptions of the phase transition between the classical magnetic phases. On the other hand, we find that the order-by-disorder mechanism makes the GSs of the pure $\Gamma$ and antiferromagnetic Kitaev models have specific orders, although the classical analyses suggest that the GSs are highly degenerate.

In summary, we carry out a comprehensive study of the $J$-$K$-$\Gamma$ model on a triangular lattice in the full parameter space and map out its global phase diagram by use of a combination of the exact diagonalization, the classical Monte Carlo simulation and analytical analyses. We find that there are five quantum phases in the limit of $\Gamma=0$.
Among them, the 120$^\circ$ N\'{e}el, the dual N\'{e}el and one of the stripe phases extend into the region with $\Gamma\ne 0$. However, the other stripe and the ferromagnetic phases are unstable in response to an infinitesimal $\Gamma$ interaction. Due to the introduction of the $\Gamma$ term, five new phases emerge including two ferromagnetic phases, one stripe, one modulated stripe and a possible quantum spin liquid. We also elaborate that the pure $\Gamma$ model has a ferromagnetic ground state and the antiferromagnetic Kitaev model a stripe ground state, which are selected by the order-by-disorder mechanism from the degenerate classical ground states.

\begin{acknowledgments}
This work was supported by the National Natural Science Foundation of China (Grants No. 11674158, No. 11774152 and No. 11774300) and National Key Projects for Research and Development of China (Grant No. 2016YFA0300401).
\end{acknowledgments}

\appendix

\section{\label{apx:PhaseBoundary}Phase boundaries in the quantum phase diagram}

\begin{figure}
    \centering
    \includegraphics[width=\columnwidth]{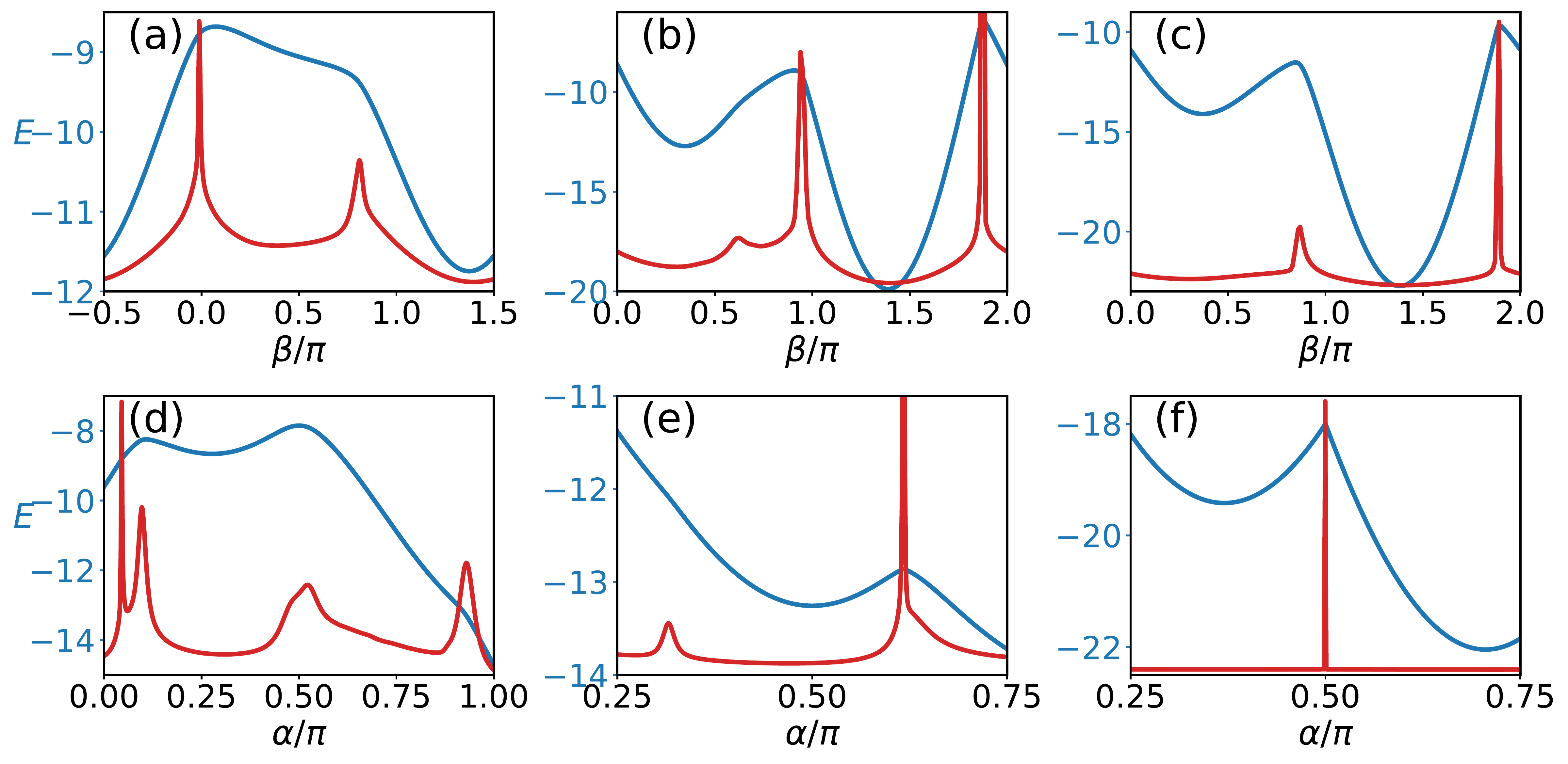}
    \caption{\label{fig:SecondDerivatives} GS energies and their second derivatives versus $\alpha$ or $\beta$ for six representative path in the phase diagram. The blue lines are the GS energies and the red lines are the second derivatives. In (a)-(c) $\alpha$ is fixed to $0.05\pi$, $0.3\pi$ and $0.75\pi$, respectively. In (d)-(f), $\beta$ is fixed to $0$, $0.5\pi$ and $1.5\pi$, respectively.}
\end{figure}

As mentioned in the main text, the phase boundaries are determined by the location of singularities in $-\partial^{2} E_{2} / \partial \alpha^{2}$ and $-\partial^{2} E_{0} / \partial \beta^{2}$ based on the ED calculations. Some representative curves are shown in Fig.~\ref{fig:SecondDerivatives}. In Fig.~\ref{fig:SecondDerivatives}(a) for $\alpha=0.05\pi$, the second derivative of $E_0$ versus $\beta$ has two peaks located at $\beta=-0.01\pi$ and $\beta=0.81\pi$, which are the phase boundaries between the FM-B and QSL phases. As we increase $\alpha$ to $0.3\pi$, there are three singularities in the second derivatives (see Fig.~\ref{fig:SecondDerivatives}(b)). The small peak at $\beta=0.62\pi$ indicates the phase transition from modulated stripe to stripe-A, while the other two sharp peaks at $\beta=0.94\pi$ and $\beta=1.87\pi$ originate from the phase transitions from stripe-A to FM-B and from FM-B to stripe-C, respectively. Fig.~\ref{fig:SecondDerivatives}(c) show the case for $\alpha=0.75\pi$, where the two peaks located at $\beta=0.87\pi$ and $\beta=1.89\pi$ signify the phase transitions between stripe-A and FM-C. On the other hand, we can also fix $\beta$ to detect the phase transitions with varying $\alpha$ and the corresponding results are shown in Fig.~\ref{fig:SecondDerivatives}(d)-(f). In Fig.~\ref{fig:SecondDerivatives}(d), $\beta$ equals to $0$, the first two sharp peaks at $\alpha=0.045\pi$ and $\alpha=0.097\pi$ reveal the phase transitions from FM-B to QSL and further to stripe-C. The peak at about $\alpha=0.5\pi$ is due to the phase transition from stripe-C to stripe-A. The last singularity at $\alpha=0.93\pi$ marks the phase boundary between stripe-A and FM-C. In Fig.~\ref{fig:SecondDerivatives}(e), we show the phase transition between modulated stripe and 120$^\circ$ N\'{e}el as well as the phase transition from 120$^\circ$ N\'{e}el to stripe-A. Here, $\beta$ is fixed to $0.5\pi$, and the transitions occur at $\alpha=0.32\pi$ and $\alpha=0.62\pi$, respectively. Last but not least, we show there are phase transitions between these different FM phases. In Fig.~\ref{fig:SecondDerivatives}(f) where $\beta=1.5\pi$, the sharp peak at $\alpha=0.5\pi$ shows that the FM-A phase for the HK model is a cirtical point, infinitesimal $\Gamma$ interaction will cause phase transition to other FM phases which have different moment direction from FM-A phase.

\section{\label{apx:SSF}Supplements of the SSF from the ED calculations}

\begin{figure}
    \centering
    \includegraphics[width=\columnwidth]{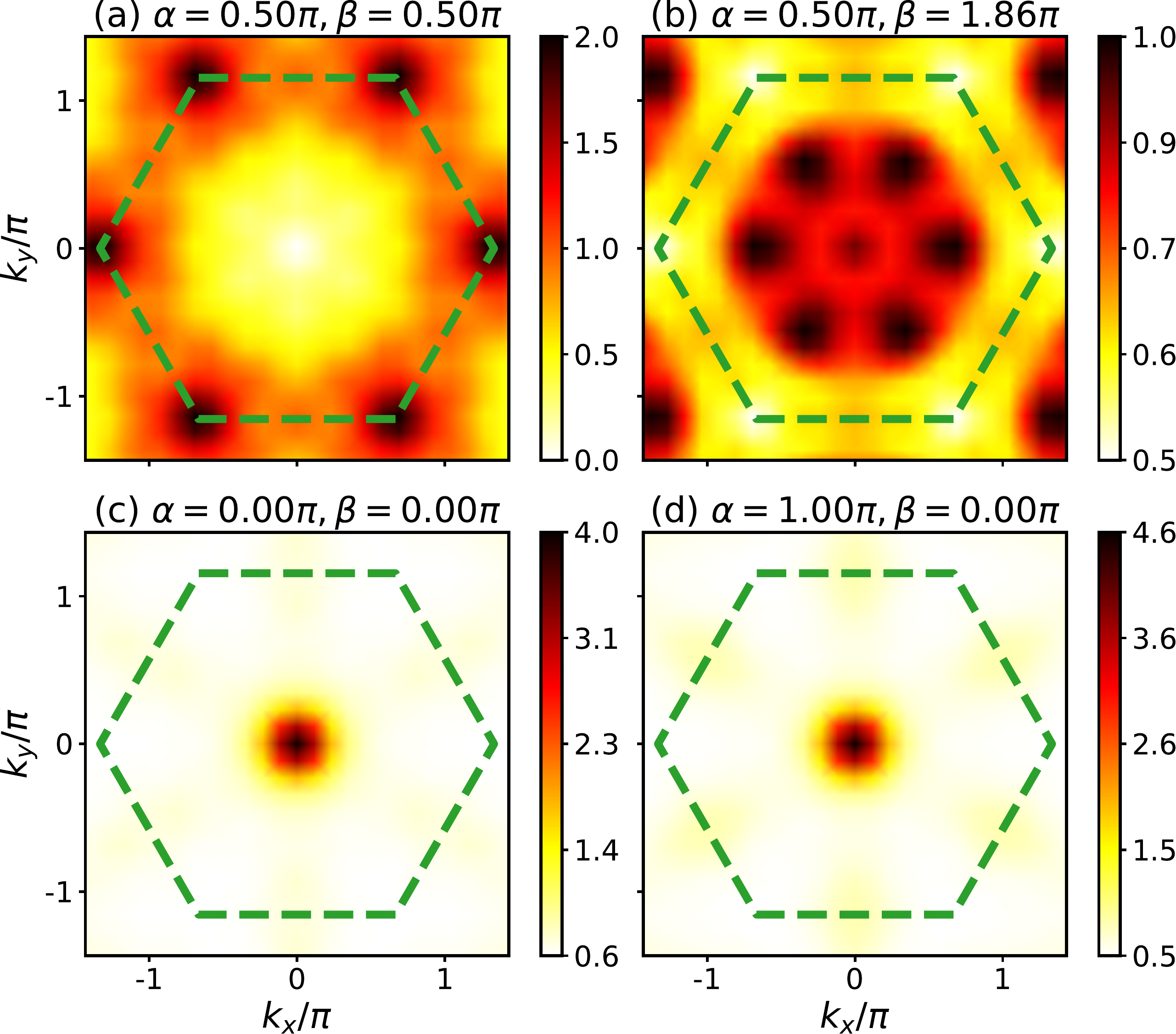}
    \caption{\label{fig:AppendixSSF}(Color online) (a) and (b) are the representative SSF profiles for 120$^\circ$ N\'{e}el and Dual N\'{e}el phase, respectively. (c) and (d) are the SSF profiles for the pure $\Gamma$ models with $\Gamma = \pm 1$, respectively.}
\end{figure}

Here we provide additional SSF profiles as a supplement to these profiles shown in the main text. For classical 120$^\circ$ N\'{e}el state, the SSF has high intensity at the corner of the first BZ, i.e., $\tilde{K}$ points and the SSF peaked at the middle points of $\tilde{\Gamma}$ and $\tilde{K}$'s for the Dual N\'{e}el state. Fig.~\ref{fig:AppendixSSF}(a) and (b) show the SSF profiles calculated from our ED ground state at $\alpha=0.5\pi, \beta=0.5\pi$ and $\alpha=0.5\pi, \beta=1.85\pi$ which located at the 120$^\circ$ N\'{e}el phase and Dual N\'{e}el phase, respectively. For both profile, the points with highest intensity are consistent with the classical analyses.

In the main text, we argue that the GSs of the pure $\Gamma$ models are FM ordered states. Here we provide the SSFs for $\Gamma = \pm 1$ in Fig.~\ref{fig:AppendixSSF}(c) and (d). Both SSFs have high intensities at the $\tilde{\Gamma}$ points, which is a typical characteristic of FM state.

\section{\label{apx:DegeneratedStates}Classical GS configurations for some special interaction parameters}

\subsection{$J=K=0$, $\Gamma=1$}

\begin{figure}
    \centering
    \includegraphics[width=\columnwidth]{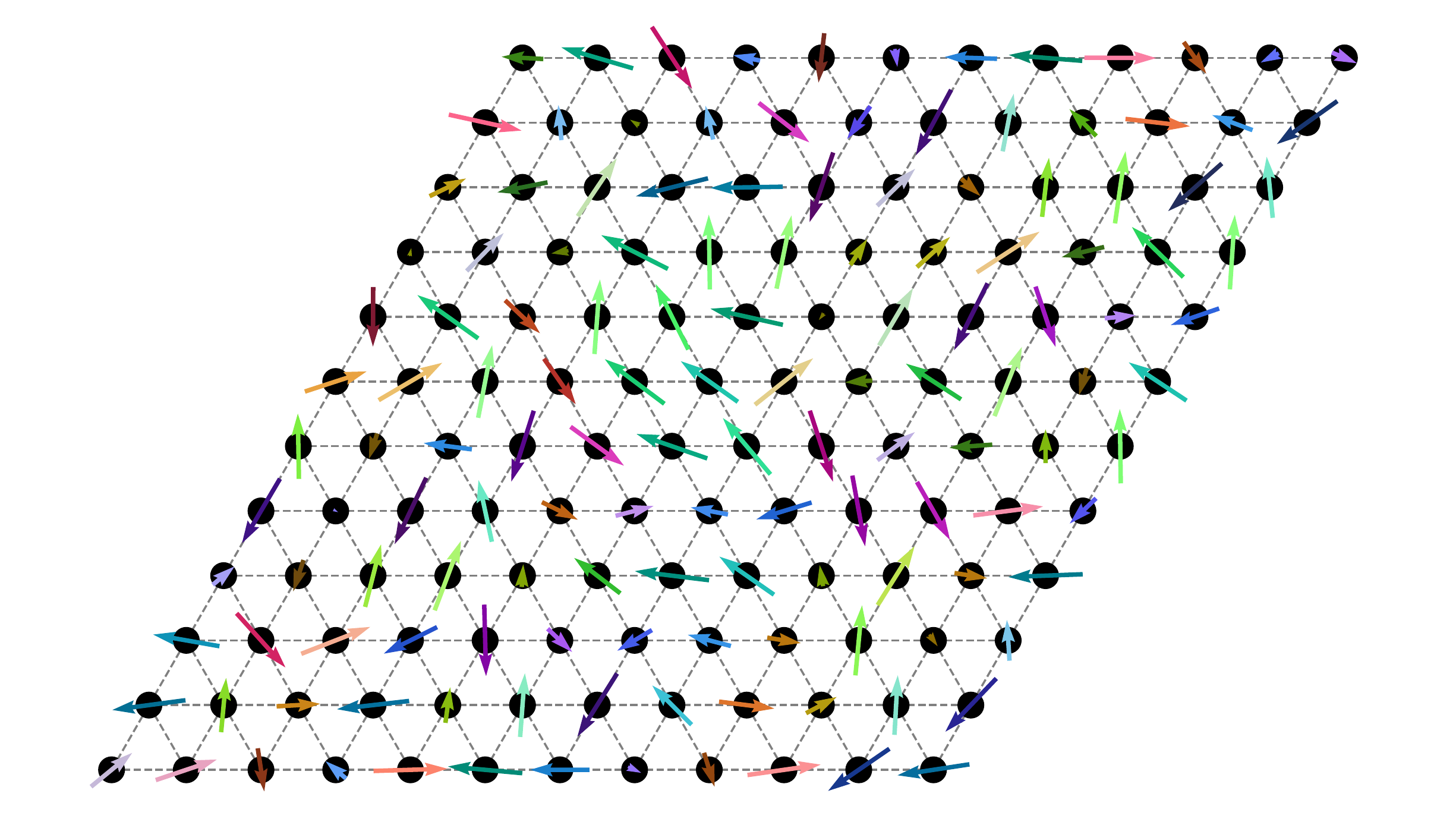}
    \caption{\label{fig:GSForPositiveGamma}(Color online) Typical disordered GS for $J=K=0, \Gamma=1$. For clarity, the equal spin vectors are marked with the same color, and the arrows denote the projections of the three-dimensional vectors to the $xy$ plane.}
\end{figure}

For pure positive $\Gamma$ model, the classical GSs are highly degenerate including FM state as well as states with no long range order. The ordered moments of the FM state lie in the lattice plane. Fig.~\ref{fig:GSForPositiveGamma} shows a typical spin configuration of those disordered states that have the same energy as the FM state.

\subsection{$J=K=0$, $\Gamma=-1$}

\begin{figure}
    \includegraphics[width=\columnwidth]{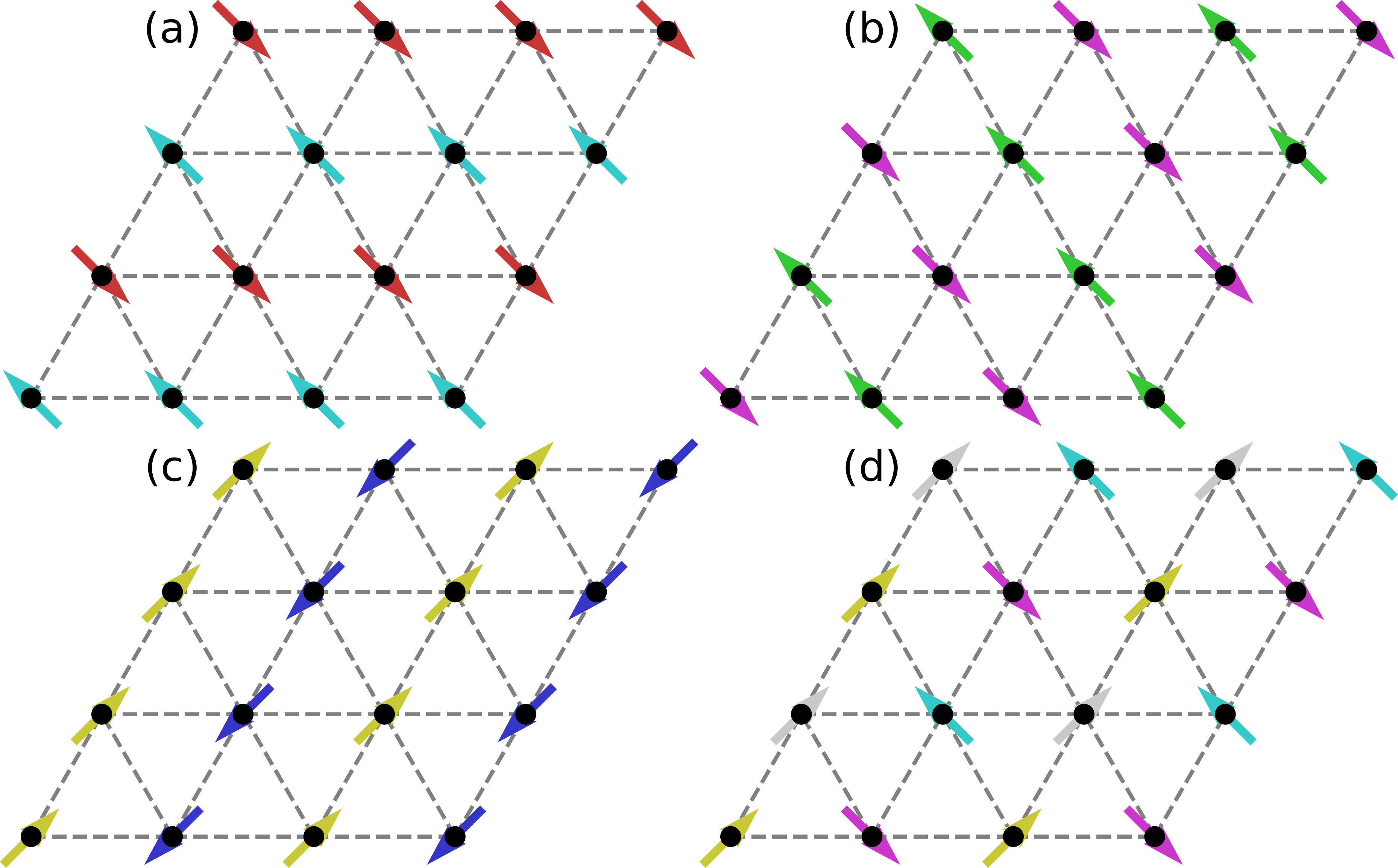}
    \caption{\label{fig:GSForNegativeGamma}(Color online) Typical classical GS spin configurations for $J=K=0, \Gamma=-1$. The spin vectors were projected to the $xy$ plane and different vectors were represented by different colors. (a)-(c) Stripe ordered states. The directions for the cyan, red, pink, green, yellow and blue arrows are $[\bar{1}11]$, $[1\bar{1}\bar{1}]$, $[1\bar{1}1]$, $[\bar{1}1\bar{1}]$, $[11\bar{1}]$ and $[\bar{1}\bar{1}1]$, respectively. (d) Noncollinear state. The directions for the yellow, gray, pink, cyan arrows are $[11\bar{1}]$, $[111]$, $[1\bar{1}1]$ and $[\bar{1}11]$, respectively.}
\end{figure}

For pure negative $\Gamma$ model, we found several energetically degenerate states as the classcial GSs, including FM state, stripe states and a noncollinear state. The ordered moment of the FM state perpendicular to the lattice plane. As for the stripe states, there are three degenerate spin configurations as shown in Fig.~\ref{fig:GSForNegativeGamma}(a)-(c) and the moment directions for the cyan, red, pink, green, yellow and blue arrows are $[\bar{1}11]$, $[1\bar{1}\bar{1}]$, $[1\bar{1}1]$, $[\bar{1}1\bar{1}]$, $[11\bar{1}]$ and $[\bar{1}\bar{1}1]$, respectively. Apart from these collinear states, a noncollinear state also exist as the classical ground state. The magnetic unit-cell contains four lattice sites, and the moment directions for the yellow, gray, pink, cyan arrows in Fig.~\ref{fig:GSForNegativeGamma}(d) are $[11\bar{1}]$, $[111]$, $[1\bar{1}1]$ and $[\bar{1}11]$, respectively.

\subsection{$J=\Gamma=0$, $K=1$}

\begin{figure}
    \includegraphics[width=\columnwidth]{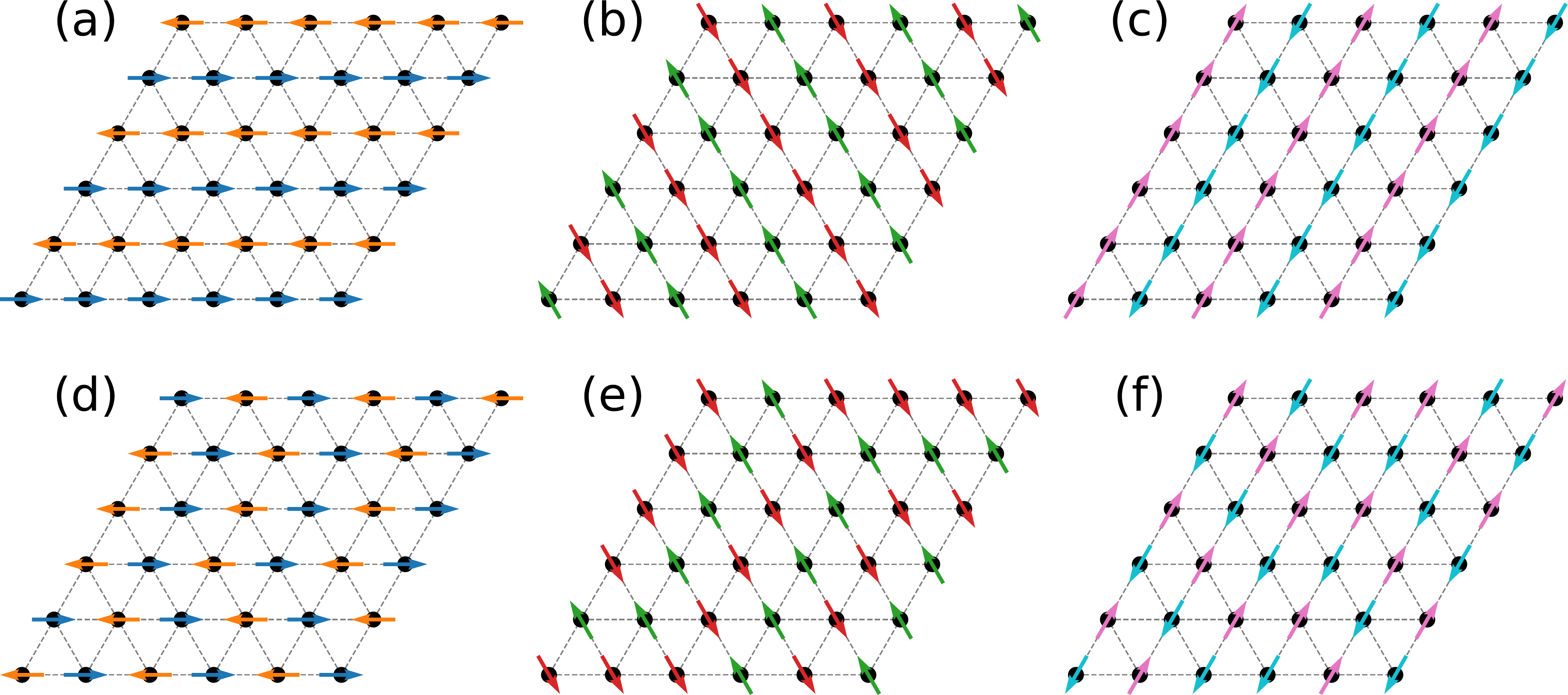}
    \caption{\label{fig:GSForPositiveK}(Color online) Classical stripe and nematic GS spin configurations for $J=\Gamma=0, K=1$. (a)-(c) Stripe states. (d)-(f) Nematic states.}
\end{figure}

The GSs for the classical antiferromagetic Kitaev model are also degenerate involving three types of nematic ordered states and three stripe ordered states. For the stripe ordered states shown in Fig.~\ref{fig:GSForPositiveK}(a)-(c), the corresponding ordered moments lie in the $yz$, $xz$ and $xy$ plane, respectively. For the nematic ordered state shown in Fig.~\ref{fig:GSForPositiveK}(d), the spins form antiferromagetic chains along the $x$-bond direction and the moment direction for the blue arrows are $[100]$. Different antiferromagnetic chains are decoupled in the nematic state. Similarly, the antiferromagetic chains along the $y$-bond and $z$-bond direction for Fig.~\ref{fig:GSForPositiveK}(e) and Fig.~\ref{fig:GSForPositiveK}(f) respectively. The direction for the green arrows in Fig.~\ref{fig:GSForPositiveK}(e) and pink arrows in Fig.~\ref{fig:GSForPositiveK}(f) are $[010]$ and $[001]$, respectively.

\bibliography{ref}

\end{document}